\documentclass[useAMS,usenatbib]{mn2e}
\usepackage{graphicx}
\usepackage{times}
\usepackage{amssymb}
\usepackage{amsmath}
\begin{document}

\title[Information Contained in BAO Measurements] 
{The Information Content of Anisotropic Baryon Acoustic Oscillation Scale Measurements}

\author[A. J. Ross et al.]{\parbox{\textwidth}{
Ashley J. Ross\thanks{Email: ross.1333@osu.edu; Ashley.Jacob.Ross@gmail.com}$^{1,2}$, 
Will J. Percival$^{2}$,
Marc Manera$^{3,2}$
}
  \vspace*{4pt} \\ 
$^{1}$Center for Cosmology and AstroParticle Physics, The Ohio State University, Columbus, OH 43210, USA\\
$^{2}$Institute of Cosmology \& Gravitation, Dennis Sciama Building, University of Portsmouth, Portsmouth, PO1 3FX, UK\\
$^{3}$University College London, Gower Street, London WC1E 6BT, UK\\
}
\date{Accepted by MNRAS} 

\pagerange{\pageref{firstpage}--\pageref{lastpage}} \pubyear{2014}
\maketitle
\label{firstpage}

\begin{abstract}
  Anisotropic measurements of the Baryon Acoustic Oscillation (BAO)
  feature within a galaxy survey enable joint inference about the
  Hubble parameter $H(z)$ and angular diameter distance
  $D_A(z)$. These measurements are typically obtained from moments of
  the measured 2-point clustering statistics, with respect to the
  cosine of the angle to the line of sight $\mu$. The position of the
  BAO features in each moment depends on a combination of $D_A(z)$ and
  $H(z)$, and measuring the positions in two or more moments breaks
  this parameter degeneracy. We derive analytic formulae for the
  parameter combinations measured from moments given by Legendre
  polynomials, power laws and top-hat Wedges in $\mu$, showing
  explicitly what is being measured by each in real-space for
  both the correlation function and power spectrum, and in
  redshift-space for the power spectrum. The large volume covered by
  modern galaxy samples means that the correlation function can be well
  approximated as having no correlations at different $\mu$ on the
  BAO scale, and that the errors on this scale are
  approximately independent of $\mu$. Using these approximations, we
  derive the information content of various moments. We show that
  measurements made using either the monopole and quadrupole, or the
  monopole and $\mu^2$ power-law moment, are optimal for anisotropic
  BAO measurements, in that they contain all of the available
  information using two moments, the minimal number required to
  measure both $H(z)$ and $D_A(z)$. We test our predictions using 600
  mock galaxy samples, matched to the SDSS-III Baryon Oscillation Spectroscopic Survey 
  CMASS sample, finding a good match to our analytic
  predictions. Our results should enable the optimal extraction of
  information from future galaxy surveys such as eBOSS, DESI and
  Euclid.
\end{abstract}
\begin{keywords}
  cosmology: observations - (cosmology:) large-scale structure of Universe
\end{keywords}

\section{Introduction}

The clustering of galaxies contains the imprint of the BAO scale, at a
fixed co-moving distance $\sim$150\,Mpc (see, e.g., \citealt{Eis05}
for a review). The apparent location of the position of the feature
along the line of sight (los) depends on the value of the Hubble
parameter, $H(z)$, and its apparent location transverse to the los
depends on the angular diameter distance, $D_A(z)$. Thus, measurements
of the clustering of galaxies along and transverse to the los allows
simultaneous measurement of $D_A(z)$ and $H(z)$ (see, e.g.,
\citealt{HuH03} and \citealt{PadW08}).

The Sloan Digital Sky Survey (SDSS; \citealt{York00}) III
\citep{Eis11} Baryon Oscillation Spectroscopic Survey (BOSS;
\citealt{Dawson12}) has provided galaxy samples large enough to
robustly measure BAO scale information along and transverse to the
los and thus independently measure $D_A(z)$ and
$H(z)$. Two methods have been applied to BOSS data that isolate the
BAO information: ``Wedges'' \citep{Kazin12,Kazin13} and ``Multipoles''
\citep{Xu13} and results using both methodologies are presented in
\cite{alph}.

As measurements become statistically more precise, there is an
increased pressure on the analysis pipeline to ensure the extraction
of information is robust. The elements of the pipeline requiring
careful consideration include the models to be fitted to the data, the
statistical procedure to be applied, accurate estimation of systematic
errors, and a precise knowledge of what is actually being measured. In
this paper, we focus on the latter issue for anisotropic BAO
measurements, considering the information content of moments of
2-point statistics. Recently, studies such as \cite{Taruya11,FB14,Blazek14} have 
also studied the information content of anisotropic clustering measurements. In our study,
we build on these results by focusing purely on the $D_A(z)$ and $H(z)$ information that can be measured via the BAO position, thereby enabling an alternative and simplified analytic treatment. Further, we focus primarily
on post-`reconstruction' clustering measurements \citep{Eis07}, where the large-scale clustering amplitudes are expected
to be isotropic. In this case, we show that moments based on polynomials of the
cosine of the angle to the los ($\mu$) are complete for any
non-degenerate set of two moments that includes zero and second order
terms. We then compare the precision of $D_A(z)$ and $H(z)$ measurements one
obtains using the Wedges and Multipoles methodology, both based on analytical predictions and empirical measurements.

Our paper is structured as follows: After developing general formulae
in Section~2, we assume that information is equally distributed with
respect to $\mu \equiv {\rm cos}(\theta_{\rm los})$, equivalent to a
spherically symmetric distribution and matching empirical results, and
in Section~3 we predict the variance and covariance expected on
$D_A(z)$ and $H(z)$ measurements for two simple combinations of
measurements: one in which a combination of the spherically averaged
clustering and clustering averaged over a $\mu^p$ window are used, and
another using Wedges split at an arbitrary $\mu_d$. In Section 4, we
describe how the BAO scale can be fitted using the different
methodologies. In Section 5, we measure the BAO scale using 600 mock
BOSS samples and compare the results obtained using each methodology,
and to the results of \cite{alph}. Where applicable, we assume the
same fiducial cosmology as in \cite{alph}: $\Omega_m = 0.274$,
$h=0.7$, $\Omega_bh^2 = 0.0224$.

\section{The Anisotropic BAO Signal}

In this section we describe our formalism for considering measurements
of the projected BAO scale including an isotropic dilation, and the
anisotropic Alcock-Paczynski effect \citep{AP}. We present our
formalism in configuration space, but our derivations are equally
valid in Fourier space and therefore applicable to $P(k,\mu)$
measurements.

The observed distance between two galaxies $r$ defined
assuming a fiducial or reference cosmological model, and the observed
cosine of the angle the pair makes with respect to the line-of-sight
(los) $\mu$ are given by
\begin{equation}
r^2 = r^2_{||}+r^2_{\perp};\,\,\,\, \mu = \frac{r_{||}}{r},
\end{equation}
where $r_{||}$ is the los separation and $r_{\perp}$ is the transverse
separation. The estimate of these separations is dependent on the
assumed cosmology. Defining
\begin{equation}
  \alpha_{||} \equiv H(z)_{\rm fid}/H(z)_{\rm true};\,\,\,\, 
  \alpha_{\perp} \equiv D_{A,{\rm true}}/D_{A,{\rm fid}},
\end{equation}
the true separation, $r^{\prime}$, is given by
\begin{equation}
  r^{\prime} = \alpha r = \sqrt{\alpha^2_{||}r^2_{||}+\alpha^2_{\perp}r^2_{\perp}}.
\end{equation}
We can re-arrange the above equations to express the stretch as a
function of the angle to the line of sight:
\begin{equation}
\label{eq:amu}
\alpha(\mu) = \sqrt{\mu^2\alpha_{||}^2+(1-\mu^2)\alpha^2_{\perp}}.
\end{equation}

Assuming symmetry around $\mu = 0$, we can consider any moment of the
2-point clustering signal as an integral over measurements made along
different directions with given $\mu$ weighting. For the correlation
function we can write
\begin{equation}  \label{eq:moment}
  \xi_F(r)=\int_0^1F(\mu)\xi(r,\mu)d\mu,
\end{equation}
where $F(\mu)$ gives the relative weight of each direction to the
moment. For the monopole of the correlation function $\xi_0(r)$, for
example, $F(\mu)=1$. In this paper we only consider functions $F(\mu)$
that are normalised, that is for which $\int_0^1F(\mu)\,d\mu=1$.

In real-space, the correlation function for galaxies in a thin slice
in $\mu$ can be written $A(\mu)\xi(r^{\prime}/\alpha(\mu))$, where
$A(\mu)$ alters the amplitude, but not the shape or BAO position. If
RSD have been removed during a ``reconstruction'' \citep{Eis07} step,
this also holds. Pre-reconstruction in redshift space, we need to
adjust the template to be fitted to allow for correlation function
shape changes \citep{Jeong14}. If $\alpha\ne1$, Eq.~(\ref{eq:moment})
describes a shift in the mean position of the BAO in the moment, which
we denote $\alpha_F$, together with a ``broadening'' of the BAO bump,
which is now the superposition of $\alpha(\mu)$, which varies as given
in Eq. \ref{eq:amu}. For cosmological models close to the fiducial
cosmology used to calculate the correlation function, the broadening
is small and is degenerate with the non-linear BAO
damping. Consequently information from the BAO feature width is
commonly neglected, with the primary measurement being the BAO
position $\alpha_F$. Information from the broadening was included in
the anisotropic BAO measurements of \citet{alph}, where models of the
moments were calculated by integrating directly over $\xi(r,\mu)$. The
additional constraints available from the observed shape of the BAO
feature mean that the contours from any single moment in
$\alpha^2_{||}$ and $\alpha^2_{\perp}$ are closed, but this closure of
the contours is not important when fitting to multiple moments, which
generally break this degeneracy much more strongly.

We seek to express the expectation for the measured stretch, $\alpha_F$, determined from a moment of the 2-point clustering signal ($\xi_F$; Eq. \ref{eq:moment}), in terms of the radial and transverse stretch through the expression for $\alpha(\mu)$ given by Eq. \ref{eq:amu}. Following the arguments above, we assume the information on $\alpha(\mu)$ is separable from the overall shape of the clustering signal. This is equivalent to the modeling used in, e.g., \cite{alph} BAO fits to the measured $P(k)$, where the model consists of a BAO feature and nuisance parameters describing the overall shape of $P(k)$, and similar to the modeling used to fit $\xi(s)$ in the same study. When this is case, the maximum likelihood $\alpha(\mu)$ determined from any measured $\xi_{meas}(\mu)$ must be independent of any other parameters. 
We further assume that information in different $\mu$ bins is independent and distributed equally (which we justify empirically in Section \ref{sec:err}), and thus the maximum likelihood stretch $\alpha_i$ in any $\mu$ bin $i$ are independent. This combination of assumptions implies that, for positive-definite windows $F(\mu)$, the maximum likelihood $\alpha_F$ obtained from $\Sigma_iF(\mu_i)\xi(\mu_i)\Delta\mu_i$ is the same as the weighted sum of individual maximum likelihood $\alpha_i$, $\Sigma_i F(\mu_i)\alpha_i\Delta\mu_i$, which is
\begin{equation}  \label{eq:a_F}
  \alpha_{F} = \int_0^1 d\mu F(\mu)\left[\mu^2\alpha^2_{||}
    +(1-\mu^2)\alpha^2_{\perp}\right]^{\frac{1}{2}}
\end{equation}
for infinitesimal bins in $\mu$ and the corresponding $\xi_F$ clustering measurements defined by Eq. \ref{eq:moment}.

One can fit to $\alpha^2(\mu)$ rather than $\alpha(\mu)$. For moments $\xi_F$, this is equivalent to measuring the weighted average of
$\alpha^2(\mu)$ over the window $F(\mu)$, whose expected maximum likelihood value we express as $\langle\alpha_F^2\rangle$ and is
somewhat simpler to interpret for some functions $F(\mu)$. In this
case, we have that
\begin{equation}  \label{eq:asq_F}
  \langle\alpha^2_{F}\rangle = \int_0^1 d\mu F(\mu)\left[\mu^2\alpha^2_{||}
    +(1-\mu^2)\alpha^2_{\perp}\right].
\end{equation}
In the following we consider both approaches, fitting for either
$\alpha_F$ or $\langle\alpha_F^2\rangle$. Note that using a
positive-definite function has the added advantage that, in real-space
or post-reconstruction, the moments have the same shape as the linear
2-point clustering measurement to first-order when
$\alpha_{||}=\alpha_{\perp}=1$. Thus they will all display a clear BAO
feature that can be easily fitted.

Any single measurement of $\alpha_F$ or $\langle\alpha_F^2\rangle$
from a moment of the correlation function or power spectrum will
result in a degenerate measurement of $\alpha_{||}$ and
$\alpha_{\perp}$. Expanding around the best-fit solution to first
order, we can fit the degeneracy direction showing that the primary
measurement of $\alpha_F$ or $\langle\alpha_F^2\rangle$ results in the
same degeneracy, with a form
\begin{equation}  \label{eq:bar_a_F}
  \alpha_{F}^{m+n} = \alpha_{||}^m \alpha_{\perp}^n,
\end{equation} 
where
\begin{align}
  m&=\left.\frac{\partial\alpha_{F}} {\partial\alpha_{||}} \right|_{\alpha_{||},\alpha_{\perp}=1}
      =\int_0^1d\mu F(\mu)\mu^2,  \label{eq:m}\\
  n&= \left.\frac{\partial\alpha_{F}}{\partial\alpha_{\perp}}\right|_{\alpha_{||},\alpha_{\perp}=1}
     =\int_0^1d\mu F(\mu)(1-\mu^2). \label{eq:n}
\end{align}
The factor $m+n$ on the left-hand side of Eq.~(\ref{eq:bar_a_F})
renormalises $\alpha_{F}$ to the correct units.

In the following we consider particular forms for the function
$F(\mu)$. The analysis should be valid for both power spectrum and
correlation function analyses.

\subsection{Fitting the monopole}

For the monopole in real-space, $F(\mu)=1$, and Eqns.~(\ref{eq:m})
\&~(\ref{eq:n}) give that $m=\frac{1}{3}$ and $n=\frac{2}{3}$, and one
recovers the well-known result that BAO fits to the monopole constrain
$\alpha_{F}=\alpha_{||}^{\frac{1}{3}}\alpha_{\perp}^{\frac{2}{3}}$,
whose corresponding distance is commonly called $D_V$. Note that, for
measurements of the dilation scale parameterised by
$\langle\alpha_F^2\rangle$, the fit constrains a linear combination of
$\alpha_{||}^2$ and $\alpha_{\perp}^2$
\begin{equation}
 \langle\alpha_F^2\rangle=\frac{1}{3}\alpha_{||}^2+\frac{2}{3}\alpha_{\perp}^2.
\end{equation}
For the monopole of the power spectrum in redshift-space, 
\begin{equation}
  F(\mu)=\frac{(1+\beta\mu^2)^2}{1+\frac{2}{3}\beta+\frac{1}{5}\beta^2},
\end{equation}
including the increase in clustering amplitude driven by the
Redshift-Space Distortions \citep{Kaiser87}. Here $\beta=f/b$, where
$f$ is the logarithmic derivative of the linear growth rate with
respect to the scale factor, and $b$ is a linear deterministic
bias. Substituting this into Eqns.~(\ref{eq:m}) \&~(\ref{eq:n}) and defining $A = 1+\frac{2}{3}\beta+\frac{1}{5}\beta^2$, gives
that
\begin{equation}
  m=\frac{1}{A}\left(\frac{1}{3}+\frac{2\beta}{5}+\frac{\beta^2}{7}\right),\,\,\,\,
  n=\frac{1}{A}\left(\frac{2}{3}+\frac{4\beta}{15}+\frac{2\beta^2}{35}\right).
\end{equation}
For the SDSS-III BOSS \citep{Dawson12} CMASS galaxies,
\citet{Samushia14} measured $\beta=0.34$, which translates to $m=0.49$
and $n=0.76$ suggesting that, to first order, the BAO-scale
constraints from the monopole power spectrum measurement depend on
$\alpha_{F}=\alpha_{||}^{0.39}\alpha_{\perp}^{0.61}$. As
expected, an increase in the clustering strength along the los leads
to an increased dependence on $\alpha_{||}$ in the resulting
measurement.

Post-reconstruction, it is standard to ``approximately remove'' the
RSD based on the estimate of the potential obtained, leaving a
clustering signal whose amplitude is approximately independent of
$\mu$ (e.g., \citealt{Pad12,burden14}). Spherical averaging to give
the monopole means that there is no $\beta$-dependent term, and the
dependence of the monopole will revert to the real-space value. Note
that in this case, or in real-space, all equations are valid for both the
correlation function and the power spectrum.

\subsection{Fitting power-law moments}
\label{sec:pfit}
The Legendre polynomials form an orthogonal basis and are the standard
approach to measuring anisotropic clustering. However, using such
bases, we can have $F(\mu)<0$ for some $\mu$, and consequently, the
recovered clustering signal cannot be considered as a sum of the
clustering signals in different directions (Eq.~\ref{eq:a_F} no longer
holds). The interpretation of these moments is therefore complicated
as the BAO information is not compressible into a single stretch value.

This is not true if we instead consider the power law moments from
which the multipoles are composed. For a power law moment of the power
spectrum in redshift-space,
\begin{equation}
  F(\mu)=\frac{\mu^p (1+\beta\mu^2)^2}{(p+1)^{-1}+2\beta(p+3)^{-1}+\beta^2(p+5)^{-1}},
\end{equation} 
and
\begin{align}
  m&=\frac{1}{p+3}+\frac{2\beta}{p+5}+\frac{\beta^2}{p+7},\\
  n &=\frac{1}{p+1}+\frac{2\beta-1}{p+3}+\frac{\beta^2-2\beta}{p+5}-\frac{\beta^2}{p+7}.
\end{align}
When $\beta=0$, or post-reconstruction with RSD removal, this reduces
to constraining
$\alpha_{F}=\alpha_{||}^{\frac{p+1}{p+3}}\alpha_{\perp}^{\frac{2}{p+3}}$,
which is valid for both the correlation function and power spectrum.

Note that fits to $\langle\alpha_F^2\rangle$ constrain a linear combination of
$\alpha_{||}^2$ and $\alpha_{\perp}^2$
\begin{equation}
  \langle\alpha_F^2\rangle=m\alpha_{||}^2+n\alpha_{\perp}^2,
  \label{eq:alphaF2mn}
\end{equation}
and a first order expansion as described above does not simplify the analysis.
The degeneracy directions for $F(\mu) = 1, 3\int_0^1d\mu\mu^2$, and $5\int_0^1d\mu\mu^4$
are displayed with black dashed curves in Fig.~\ref{fig:degencom}. As $p$ increases,
the moments depend increasingly strongly on $\alpha^2_{||}$ compared
with $\alpha^2_{\perp}$.

As the Legendre multipoles are simply linear combinations of power-law
moments, the combination of the monopole and quadrupole will contain
the same information as the combination of the monopole and the $p=2$
power-law moment. Consequently, BAO fits to either the monopole and
quadrupole or to the $\mu^0$ and $\mu^2$ moments will provide the same
information and, similarly, including either or hexadecapole and $\mu^4$ moment will add the same information.

\subsection{Fitting Wedges}
\label{sec:wfit}
One could also consider setting $F(\mu)$ to be a top-hat function in
$\mu$, for example splitting the monopole into two components
separated at $\mu_d$. Such moments have been termed `Wedges'
\citep{Kazin12,Kazin13}. Using a subscript `1' for $F(\mu)=1/\mu_d$ for
$0\le\mu\le\mu_d$, and a subscript `2' for $F(\mu)=1/(1-\mu_d)$ for
$\mu_d\le\mu\le1$, one finds in real-space that

\begin{align}
m_1 & =  \frac{\mu_d^2}{3},\,\,\,\, n_1 = 1-\frac{\mu_d^2}{3};\\
m_2 & = \frac{(1-\mu_d^3)}{3(1-\mu_d)},\,\,\,\, n_2 = \frac{(2-3\mu_d+\mu_d^3)}{3(1-\mu_d)},
\end{align}
which give the coefficients for both the approximation for
$\alpha_{F}$ of Eq.~(\ref{eq:bar_a_F}) and exact solution for
$\alpha^2_{F}$ given by Eq.~\ref{eq:alphaF2mn}.

\begin{figure}
\includegraphics[width=84mm]{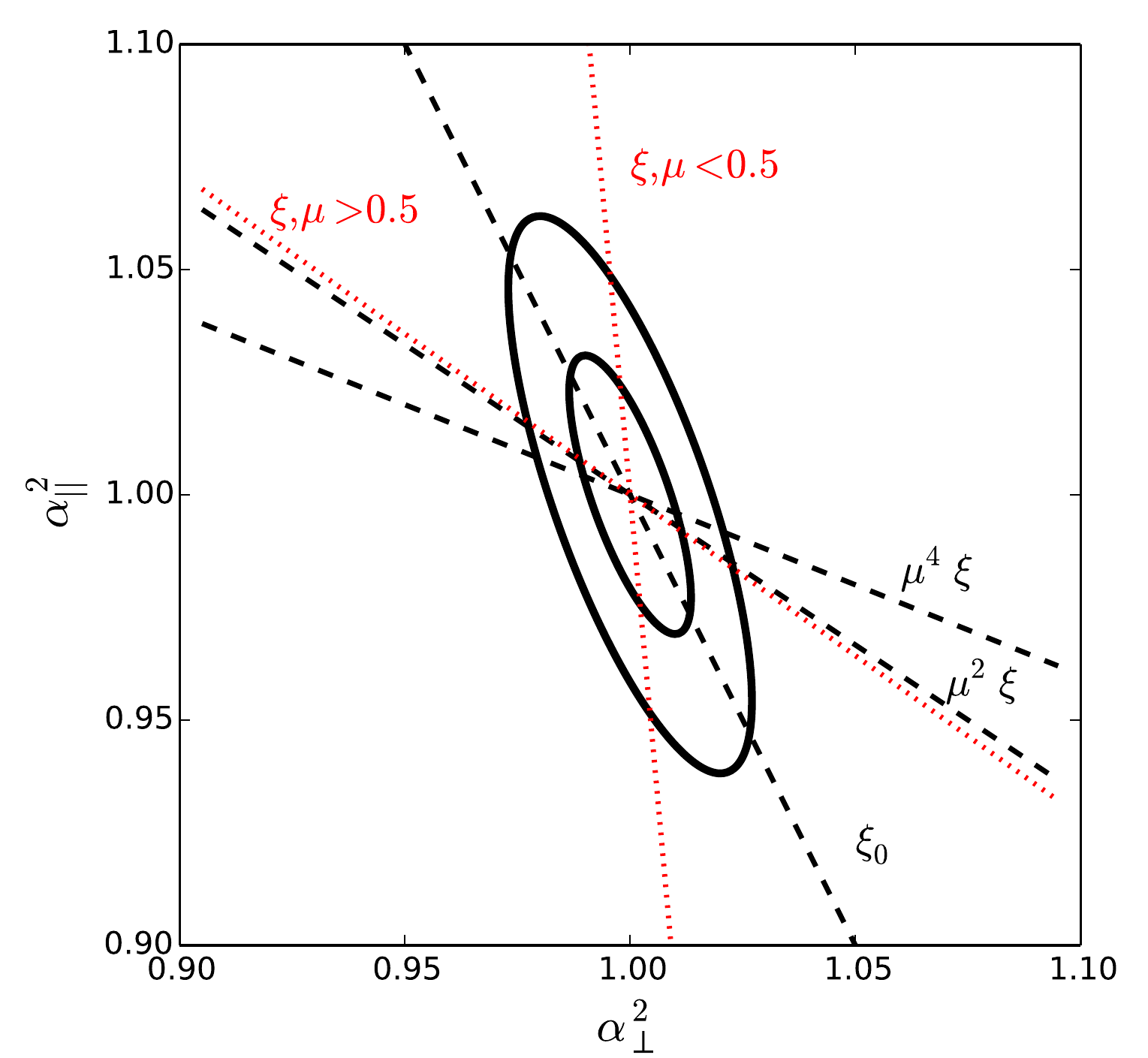}
  \caption{Straight-line curves denote the degeneracy direction between $\alpha^2_{||}$ and $\alpha^2_{\perp}$ for various (single) clustering moments. Black dashed curves denote power-law moments $\int_0^1d\mu\mu^p\xi(\mu)/\int_0^1d\mu\mu^p$, which we denote $\mu^p\xi$, and red dotted curves denote Wedges. The solid ellipses denote the 1 and 2$\sigma$ contours for the optimal combination of two moments, as derived in the following section.}
  \label{fig:degencom}
\end{figure}

Fig.~\ref{fig:degencom} displays the degeneracy directions between $\alpha^2_{||}$ and $\alpha^2_{\perp}$ for the two wedges split at $\mu_d=0.5$ using red dotted curves. The wedge with $\mu < 0.5$ constrains $\alpha^2_{\perp}$ almost exclusively and the $\mu > 0.5$ moment has a similar degeneracy as the $\mu^2$ power-law moment.

\subsection{Fitting the Quadrupole}
\label{sec:qfit}
While the idea of measuring an average BAO position does not work with
more general $F(\mu)$ models, the primary source of signal from the
quadrupole is the strength of a feature proportional to the derivative of $\xi_0$
(see, e.g., \citealt{PadW08,Xu13}). Therefore, in real-space, where there is no RSD, the
amplitude of the BAO feature observed in the quadrupole carries the majority of the information on 
 $\alpha_{||}$ and $\alpha_{\perp}$ (as opposed to any other characteristic of the quadrupole). The amplitude of the quadrupole, relative to the underlying correlation function, depends on
$\alpha_{||}$ and $\alpha_{\perp}$ through
\begin{align}
  \frac{1}{\xi(r)}
    \left.\frac{\partial\xi_2(\alpha r)}{\partial\alpha_{||}}\right|_{\alpha_{||},\alpha_{\perp}=1}
    &= \frac{\partial\log\xi(r)}{\partial\log r}
    \int_0^1\mu^2(3\mu^2-1)d\mu,  \label{eq:xi2_par} \\
  \frac{1}{\xi(r)}
  \left.\frac{\partial\xi_2(\alpha r)}{\partial\alpha_{\perp}}\right|_{\alpha_{||},\alpha_{\perp}=1}
    &= \frac{\partial\log\xi(r)}{\partial\log r}
    \int_0^1(1-\mu^2)(3\mu^2-1)d\mu \label{eq:xi2_perp}.
\end{align}
The integrals in Eqns.~(\ref{eq:xi2_par}) \&~(\ref{eq:xi2_perp}) reduce
to $\frac{4}{15}$ and $-\frac{4}{15}$ respectively, showing that the
dependence on $\alpha_{||}$ and $\alpha_{\perp}$ is equal and
opposite, suggesting that the measurement will constrain
\begin{equation}
  \frac{\alpha_{||}}{\alpha_\perp}
    \frac{\partial\log\xi(r)}{\partial\log r},
\end{equation}
to first order, matching the dominant term in the expansion of \citet{Xu13}.

\section{Errors on measured moments}
\label{sec:err}

If we can model the distribution of signal-to-noise of modes as a
function of $\mu$, we can predict the possible constraints one may
obtain on $\alpha_{||}$ and $\alpha_{\perp}$. In redshift-space, on
large-scales the modes have signal-to-noise that varies with $\mu$,
with the linear $(1+\beta\mu^2)^2$ term increasing the amplitude of
the power spectrum, which reduces the impact of the shot noise along
the los. Although the amplitude of the modes are usually renormalised
with the removal of the RSD during the reconstruction process, the
signal-to-noise remains $\mu$-dependent, as the ``RSD removal'' is
effectively a renormalisation of the redshift-space modes, rather than
a removal of signal \citep{burden14}. 

\begin{table}
\caption{BAO measurements on mocks as a function of $\mu$. $\langle \alpha \rangle$ is the mean recovered stretch parameter (the relative BAO scale in that $\mu$ window), $\langle \sigma \rangle$ is the mean recovered uncertainty on $\alpha$, and $S$ is the standard deviation of the recovered $\alpha$. }
\begin{tabular}{lcccc}
\hline
\hline
$\mu$ range  &  $\langle \alpha \rangle$ & $\langle \sigma \rangle$ & $S$ & \#\\
\hline
$0<\mu<0.2$ & 0.998 & 0.022 & 0.021 & 0 \\
$0.2<\mu<0.4$ & 1.000 & 0.021 & 0.021 & 1 \\
$0.4<\mu<0.6$ & 0.999 & 0.019 & 0.019 & 2 \\
$0.6<\mu<0.8$ & 1.001 & 0.021 & 0.020 & 3 \\
$0.8<\mu<1.0$ & 1.003 & 0.022 & 0.021 & 4 \\
\hline
\label{tab:mu-noise}
\end{tabular}
\end{table}

The window function will also affect the signal-to-noise as a function
of $\mu$ in the correlation function by varying the pair numbers, and
in the power spectrum by reducing the number of independent
modes. However, for samples such as BOSS CMASS, the window has a
negligible effect, and the statistical distribution of pairs is close
to being isotropic except on very large scales. On small scales, the
BAO damping is asymmetric, and radial effects such as the
Fingers-of-God (FoG) become important. Thus, we might expect the
distribution of signal-to-noise to be a complicated function of
$\mu$.

We investigate the amount of BAO information as a function of $\mu$ empirically, using the methods
described in Section~\ref{sec:baofit}, and the post-reconstruction
mock catalogues for the BOSS CMASS sample, described in
Section~\ref{sec:BAOmeas}. We split the data into broad bins in $\mu$ and find the mean uncertainty and variance for BAO measurements in these bins. We present this information in Table~\ref{tab:mu-noise}, which shows that the BAO information is close to having an even distribution in $\mu$ for the correlation function. Minima for the recovered uncertainty and
standard deviation on the measured BAO scale are found in the $0.4 <
\mu < 0.6$ bin. A potential explanation is that between $0 < \mu < 0.5$ the effects of linear RSD boost the BAO signal, but at larger $\mu$ effects such as FoG remove information and reduce the signal-to-noise. Regardless, this minimum is shallow: the difference in recovered uncertainty is at most 15 per cent and the results therefore justify our choice to treat the information as constant in $\mu$.

\begin{figure}
\includegraphics[width=84mm]{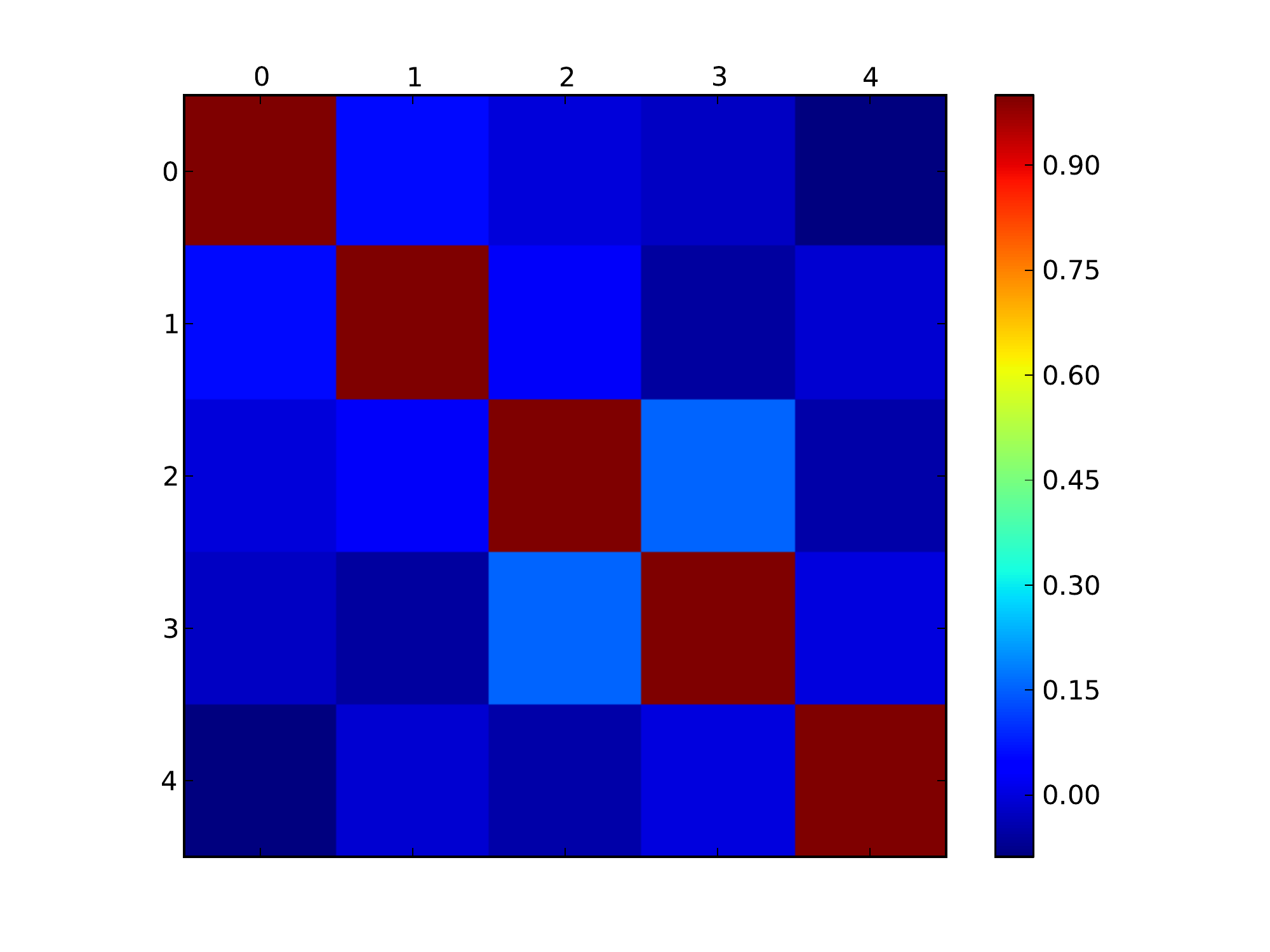}
\caption{The correlation matrix of the five BAO measurements made in
  0.2 thick $\mu$ bins obtained from 600 BOSS CMASS mocks, as
  described in Table \ref{tab:mu-noise} (and numbered in the same
  manner).}
  \label{fig:baocorr}
\end{figure}

One may also worry about correlation between the clustering at
different $\mu$. For the power spectrum and an infinite volume, one
expects no correlation between the clustering measured at different
$\mu$. Once a survey window is applied, correlations will be induced,
but for a survey the size of BOSS we expect these correlations to be
small at the BAO scale. We measure the correlation between the BAO
measurements in the five $\mu$ bins described in
Table~\ref{tab:mu-noise} and we display the correlation matrix in
Fig.~\ref{fig:baocorr}. We find the magnitude of correlations is at
most 0.15, and we expect the power spectrum to be significantly less
correlated than the correlation function. We therefore ignore any correlations between the BAO information at different
$\mu$ in our analytical derivations that follow.

The results we presented in this section suggest that, to a good approximation, one can treat the distribution in $\mu$ of BAO information in the post-reconstruction DR11 BOSS CMASS sample as the same as that of an infinite real-space volume. However, the distribution for any given survey may vary based on
the particular survey geometry, satellite velocities of the galaxy population 
(which smear the BAO feature at high $\mu$), and the magnitude of the boost in clustering amplitude due
to linear RSD effects (which boosts the high $\mu$ signal).

\subsection{Complete sets of estimators}  \label{sec:likelihood}

Suppose that we have measured $\alpha^2_{\rm meas,\mu}$ in a series of
(independent) bins in $\mu$ (which we can treat as infinite in
number), then fitting these measurements with parameters
$\alpha^2_{||}$ and $\alpha^2_{\perp}$ would minimise
\begin{equation}  \label{eq:chisq}
  \chi^2 = \int_0^1\,d\mu\,
    \sigma_0^{-2}\left[\mu^2\alpha^2_{||}+(1-\mu^2)\alpha^2_{\perp}
    -\alpha^2_{\rm meas,\mu}\right]^2,
\end{equation}
where we have assumed that the value of $\alpha^2_{\rm meas,\mu}$ at a
particular $\mu$ can be represented by a Gaussian random variable with
expectation 0 and total variance $\sigma^2_0$ across all
$\mu$. Furthermore, we have assumed that the noise is evenly
distributed in $\mu$. 

The maximum likelihood estimator for
$(\alpha^2_{||},\alpha^2_{\perp})$ can be calculated by finding the
$\chi^2$ minima, solving the equations ${\bf \nabla\chi^2}={\bf 0}$,
where
\begin{equation}  \label{eq:MLE}
  {\bf \nabla\chi^2}=\frac{1}{15}
  \left(\begin{array}{cc} 3 & 2 \\ 2 & 8 \end{array}\right)
  \left(\begin{array}{cc} \alpha^2_{||}  \\ \alpha^2_{\perp} \end{array}\right) 
  - \left(\begin{array}{cc}
        \int_0^1\,d\mu\,\mu^2\alpha^2_{\rm meas,\mu}  \\
        \int_0^1\,d\mu\,(1-\mu^2)\alpha^2_{\rm meas,\mu} 
  \end{array}\right).
\end{equation}
Following Eq.~(\ref{eq:asq_F}), the measured value
$\int_0^1\,d\mu\,\mu^2 \alpha^2_{\rm meas,\mu}$ is a linear transform of that
recovered from a moment of the 2-point function with $F(\mu)=3\mu^2$,
and similarly for the $F(\mu)=(1-3\mu^2)$ moment. Looking at both
``measurements'', we see that the maximum likelihood points are fully
determined by the $p=0$ and $p=2$ power law moments, or equivalently
by the monopole and quadrupole. Note that Eq.~(\ref{eq:MLE}) relies on
the fact that the model linearly depends on the parameters
$(\alpha^2_{||},\alpha^2_{\perp})$, and does not hold, for example,
for fits to $(\alpha_{||},\alpha_{\perp})$.

Eq.~(\ref{eq:chisq}) can also be
written in terms of $(\alpha^2_{||},\alpha^2_{\perp})^T$, with an
inverse covariance matrix given by
\begin{equation}  \label{eq:Cinv}
  C^{-1}_{\alpha^2_{||},\alpha^2_{\perp}}=\frac{1}{15\sigma_0^2}
  \left(\begin{array}{cc} 3 & 2 \\ 2 & 8 \end{array}\right).
\end{equation}
This can be calculated from the second derivatives of
Eq.~(\ref{eq:chisq}). 

\subsection{Predicted errors}

In Section~\ref{sec:likelihood}, we saw how the likelihood can be
manipulated to understand constraints on $\alpha^2_{||}$ and
$\alpha^2_{\perp}$ from complete information (so the likelihood can be
rewritten in terms of the new statistics), or the two $p=0$ and $p=2$
moments of those measurements. For fits to $\alpha_{||}$ and
$\alpha_{\perp}$ or for different moments, the likelihood derived is
no longer complete. Instead, we recognize that for more general moments of positive-definite functions $F_1(\mu)$,
$F_2(\mu)$ the covariance matrix is given by\footnote{This is the general formula for covariance between the means of two Gaussian random variables with arbitrary $F(\mu)$ weighting and variance $\sigma_0^2$ for $F(\mu) = 1$. It does not depend on the definition of $\alpha$ or $\mu$.}
\begin{equation}
  \sigma_{1,2} = \sigma^2_0\int_0^1d\mu F_1(\mu)F_2(\mu),
  \label{eq:precov}
  \end{equation} 
and we use this formula throughout this section to determine the expected uncertainty on and covariance between $\alpha_{||}$ and $\alpha_{\perp}$ when using clustering measurements for pairs of $F(\mu)$ windows.

For a general power law moment, $F(\mu)=(1+p)\mu^p$, Eq. \ref{eq:precov} yields
$\sigma^2_p = \frac{(p+1)^2}{1+2p}\sigma^2_0$. The covariance between
an isotropic weighting and an arbitrary one is $\sigma_{0,F} =
\sigma^2_0$. This implies that, in our formulation, introducing a measurement over
a second window in $\mu$ as well as the monopole, does not provide
extra information on the total stretch, it only provides a way to
determine the radial and transverse components of the stretch.

\begin{figure}
\includegraphics[width=84mm]{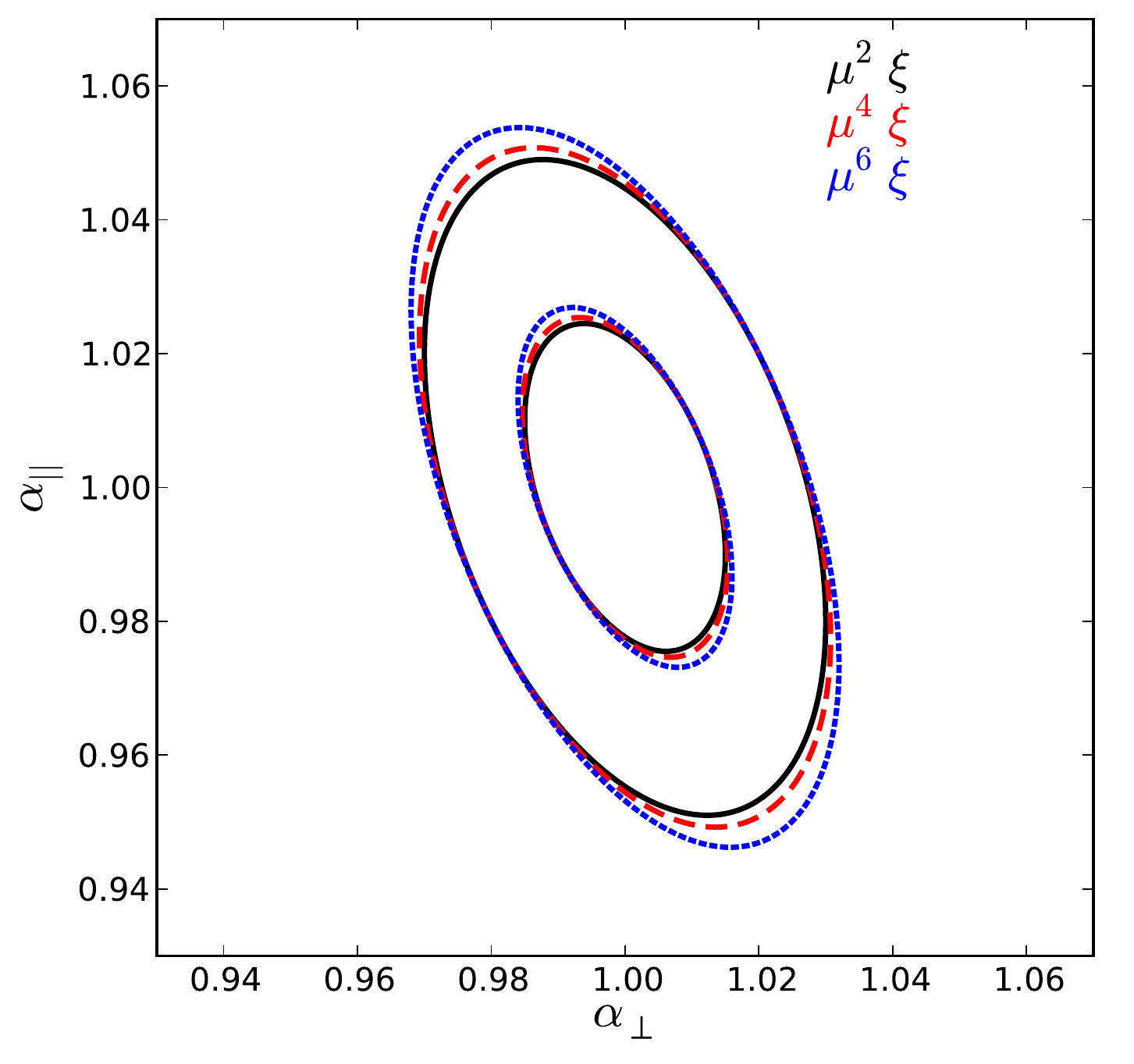}
  \caption{Ellipses showing the 1 and 2$\sigma$ contours for $\alpha_{||}$ and $\alpha_{\perp}$, expected when combining the monopole with a power-law weighted moment $\int_0^1d\mu\mu^p\xi(\mu)/\int_0^1d\mu\mu^p$, which we denote $\mu^p\xi$, for $p =2,4,6$ (black, red, blue).}
  \label{fig:ellipsecomp}
\end{figure}

Assuming a combination of measurements for $p_1=0, p_2=p$, the radial
and transverse stretch are given by (see Eqs.~\ref{eq:m} and~\ref{eq:n})
\begin{equation}
\alpha_{\perp} = \left(\alpha_0^{(3p+3)}\alpha_p^{-(p+3)}\right)^{\frac{1}{2p}}, \alpha_{||} = \left(\alpha_0^{-3}\alpha_p^{p+3}\right)^{\frac{1}{p}}
\end{equation}
and we obtain the expected uncertainty on $\alpha_{\perp}, \alpha_{||}$
\begin{equation}
\sigma^2_{||} = \sigma^2_0\frac{p^2+8p+10}{1+2p},
\label{eq:sn1}
\end{equation}
\begin{equation}
\sigma^2_{\perp} = \sigma^2_0\frac{(p+13)(p+1)}{8p+4},
\label{eq:sn2}
\end{equation}
and covariance $\sigma_{||,\perp}$
\begin{equation}
\sigma_{||,\perp} = -\sigma_0^2\frac{p^2+2p+7}{4p+2}.
\end{equation} 
For $p=2$, these equations reduce to the inverse of the matrix in
Eq.~(\ref{eq:Cinv}). The variance and the correlation, $C_{||,\perp} =
\sigma_{||,\perp}/(\sigma_{||}\sigma_{\perp})$, are minimised for $p =
2$. Inspection of these results further reveals that they match those
recovered in Section \ref{sec:likelihood} for the optimal
solution. Thus, we recover the same results using these approximate
formulae as recovered for the (not approximate) ML solutions to
measurements of $\alpha_F^2$, in the case where the ML solution is
tested. We illustrate these results by plotting the 1 and 2$\sigma$
contours predicted by these sets of covariances for $p = 2$ (black,
solid), 4 (red, dashed), and 6 (blue, dotted) in
Fig.~\ref{fig:ellipsecomp}. The length of the minor axis stays nearly
constant; which is in a similar direction to the measurement from the
monopole (see Fig.~\ref{fig:degencom}). 

For Wedges, Eq.~(\ref{eq:precov}) yields $\sigma_{1,2} = 0$ and $\sigma_1^2 = \sigma_0^2/\mu_d$, $\sigma_2^2 = \sigma_0^2/(1-\mu_d)$. Given zero correlation between non-overlapping Wedges, in principle one may gain information by using an arbitrarily large number of (non-overlapping) Wedges. However, we have shown that just two moments, equivalent to the monopole and quadrupole, form a complete set estimators. Thus, we investigate only the case where two, non-overlapping, Wedges are used\footnote{In the limit of infinite Wedges, the predicted uncertainties will clearly converge to that of the monopole and quadrupole}. We predict the uncertainties and covariance as a function of the Wedge split, $\mu_d$, to be
\begin{equation}
\sigma_{||}^2 = \sigma_0^2\left(\frac{1}{\mu_d}\left[\frac{2\mu_d^2+\mu_d^5-3\mu_d^3}{\mu_d^2(\mu_d^2-1)}\right]^2+\frac{1}{1-\mu_d}\left[\frac{3-\mu_d^2}{\mu_d+1}\right]^2\right),
\label{eq:sw1}
\end{equation}

\begin{equation}
\sigma_{\perp}^2 = \sigma_0^2\left(\frac{1}{\mu_d}\left[\frac{\mu_d^3-1}{\mu_d^2-1}\right]^2+\frac{1}{1-\mu_d}\left[\frac{\mu_d^2}{\mu_d+1}\right]^2\right),
\label{eq:sw2}
\end{equation}
and
\begin{equation}
\sigma_{||,\perp} = \sigma_0^2\left(\frac{(2+\mu_d^3-3\mu_d)(\mu_d^3-1)}{\mu_d(\mu_d-1)^2}-\frac{\mu_d^2(3-\mu_d^2)}{(1-\mu_d)(\mu_d+1)^2}\right).
\end{equation}

\begin{figure}
\includegraphics[width=84mm]{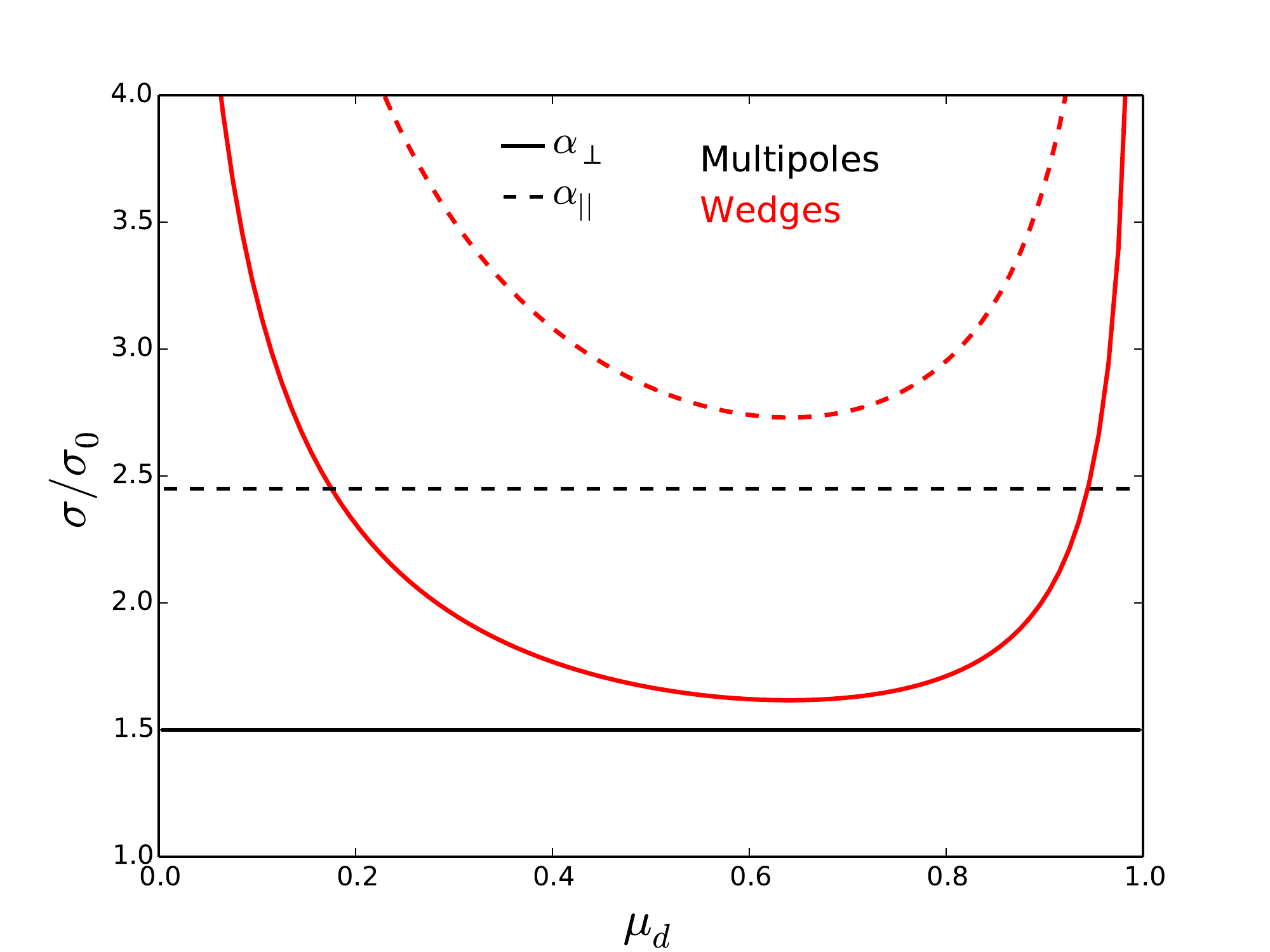}
  \caption{Red curves display the predicted uncertainty in $\alpha_{||}$ (dashed) and $\alpha_{\perp}$ (solid) recovered using Wedges, as a function of the split in $\mu$. The black curves display the predicted uncertainty for the combination of either the monopole and quadrapole, or the monopole and a $\mu^2$ window.}
  \label{fig:sigcom1}
\end{figure}

\begin{figure}
\includegraphics[width=84mm]{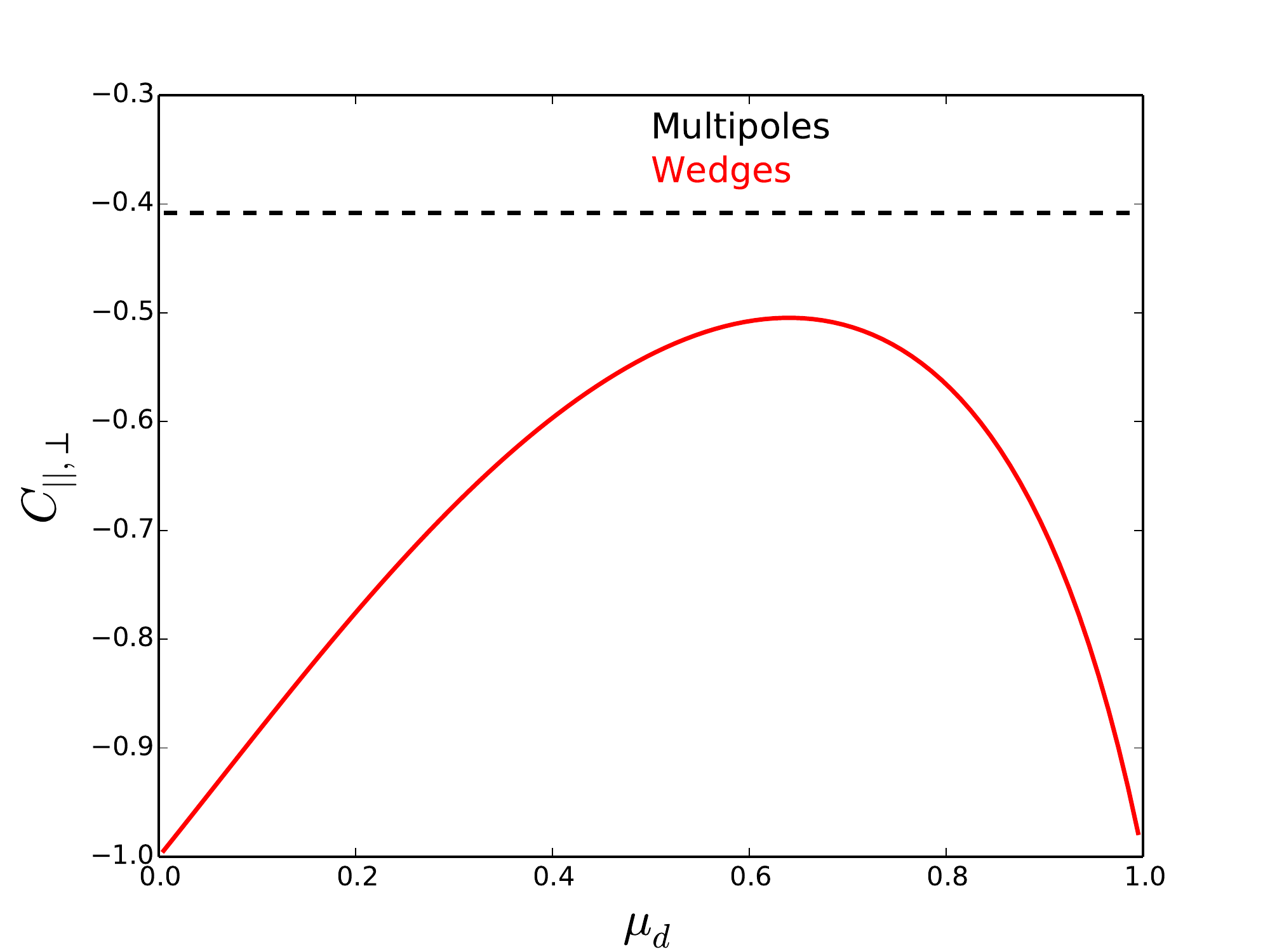}
  \caption{The solid red curve displays the predicted correlation between $\alpha_{||}$ and $\alpha_{\perp}$ recovered using Wedges, as a function of the split in $\mu$. The dashed black curve displays the predicted uncertainty for the combination of either the monopole and quadrupole, or the monopole and a $\mu^2$ window.}
  \label{fig:sigcom2}
\end{figure}

We evaluate Eqs.~(\ref{eq:sw1}) and~(\ref{eq:sw2}) for $0 < \mu_d < 1$ and compare the results to those recovered from the combination of $\alpha_0, \alpha_2$ (equivalent to the information in the monopole and quadrupole). The results are shown in Fig.~\ref{fig:sigcom1}. One can see that variance is minimised at $\mu_d = 0.64$, but that the $\alpha_0, \alpha_2$ combination always performs better. We display similar information in Fig.~\ref{fig:sigcom2}, except that we now plot the correlation $C_{||,\perp}$. Its magnitude is also minimized at $\mu_d = 0.64$ and is always greater than that of the $\alpha_0, \alpha_2$ combination.

\begin{table}
\caption{The predicted uncertainty on the radial and transverse stretch, $\sigma_{||}$ and $\sigma_{\perp}$, relative to the uncertainty on the spherically averaged stretch, and their correlation, $C_{||,\perp}$. $S$ denotes the standard deviation recovered from BAO fits to the mocks. $S_{||,\perp}$ denotes the correlation as recovered from the scatter of the BAO fits to the mocks. `W' represents Wedges, and `M' denotes the usage of $\xi_0,\xi_2$.  Compared to our predictions,  the fits to the mocks are less precise but the overall trends agree. We discuss this further in subsequent sections. }
\begin{tabular}{lcccccc}
\hline
\hline
Method  &  $\sigma_{||}$ & $\sigma_{\perp}$ &  $C_{||,\perp}$ &  $S_{||}$ & $S_{\perp}$ &  $S_{||,\perp}$\\
\hline
M & 2.45$\sigma_0$ & 1.50$\sigma_0$ & -0.41 & 2.79$\sigma_0$ & 1.58$\sigma_0$ & -0.49\\
W, $\mu_d = 0.5$ & 2.85$\sigma_0$ & 1.67$\sigma_0$ &  -0.54 & 2.98$\sigma_0$ & 1.73$\sigma_0$ & -0.56\\
W, $\mu_d = 0.64$ & 2.73$\sigma_0$ & 1.62$\sigma_0$ &  -0.50 & 3.00$\sigma_0$ & 1.66$\sigma_0$ & -0.54\\
\hline
\label{tab:pre}
\end{tabular}
\end{table}

\begin{figure}
\includegraphics[width=84mm]{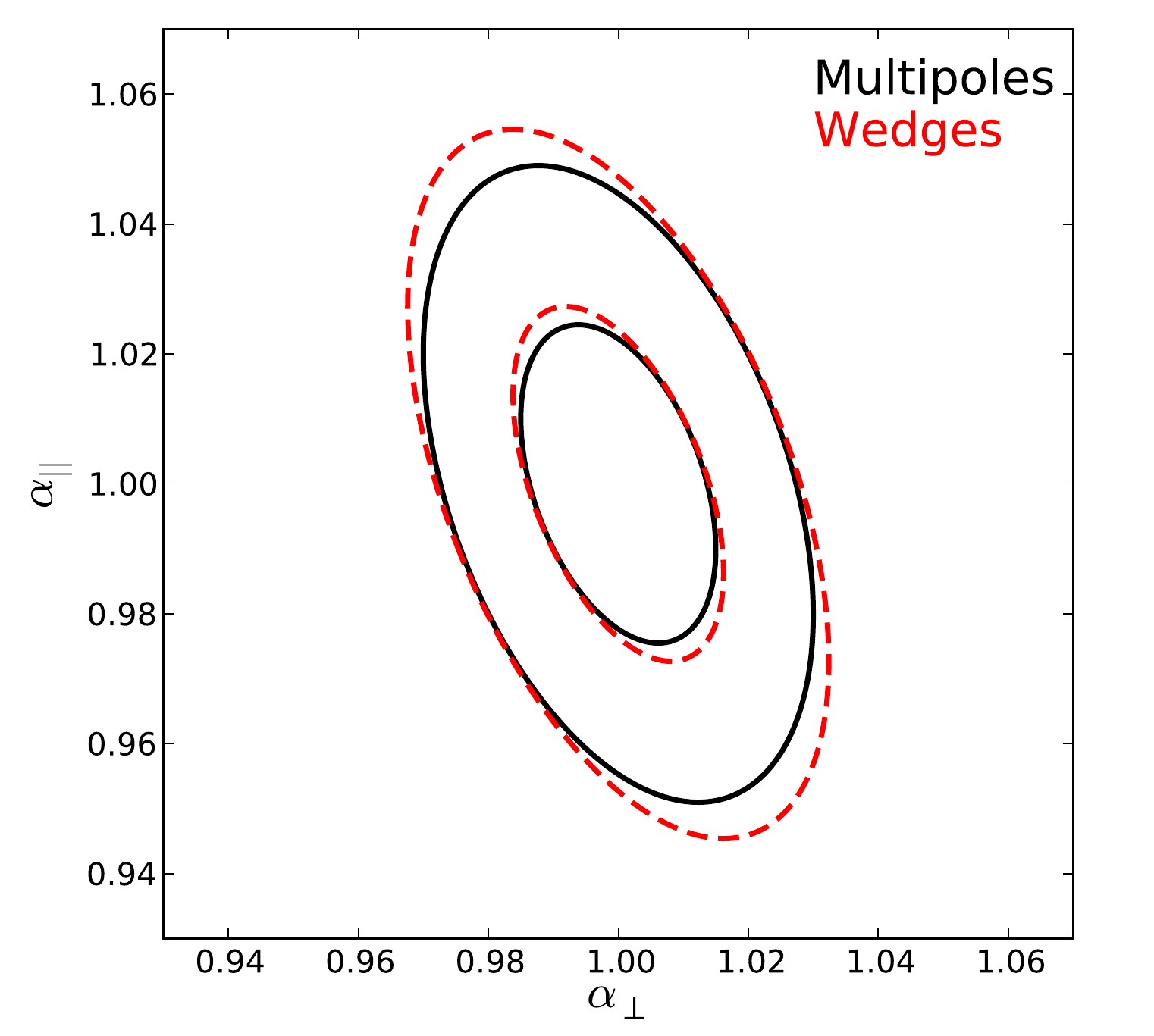}
  \caption{Ellipses showing the 1 and 2$\sigma$ contours for $\alpha_{||}$ and $\alpha_{\perp}$, expected when using multipoles (or $\mu^0$ and $\mu^2$ power-law moments; black) and when using Wedges split at $\mu_d = 0.64$.}
  \label{fig:ellipsecom}
\end{figure}

Table \ref{tab:pre} summarises the predictions we make for the recovered uncertainty on $\alpha_{||}$, $\alpha_{\perp}$ and its correlation. One can see that the predicted uncertainties on $\alpha_{||}$ and $\alpha_{\perp}$ and their covariance are worse, by close to 10 per cent for each, for Wedges than for the combination of $\xi_0$ and $\xi_2$. We illustrate this same information in Fig.~\ref{fig:ellipsecom}, where the expected 1 and 2$\sigma$ contours are displayed for Multipoles (black, solid) and Wedges split at $\mu_d = 0.64$ (red, dashed) are displayed. The major-axes of the ellipses are nearly aligned and it is along this direction that Wedges provide less-optimal constraints.

\section{BAO Fitting}
\label{sec:baofit}
We use the same model to fit anisotropic BAO scale information as applied in \cite{alph}. We use only post-reconstruction data and match all fiducial parameter choices to those used in \cite{alph}. We generate template $\xi(s)$ using the linear $P_{\rm lin}(k)$ obtained from {\sc Camb} using the same cosmology as \cite{alph} (flat, $\Omega_m = 0.274$, $\Omega_bh^2 = 0.0224$, $h=0.7$). We account for redshift-space distortion (RSD) and non-linear effects via
\begin{equation}
P(k,\mu) = F(k,\mu,\Sigma_s)\left(P_{\rm lin}e^{-k^2\sigma_v^2}+A_{MC}P_{MC}(k)\right),
\end{equation}
where
\begin{equation}
F(k,\mu,\Sigma_s) = \frac{1}{(1+k^2\mu^2\Sigma^2_s/2)^2}
\end{equation}
and
\begin{equation}
P_{MC}(k) = 2 \int\frac{d^3q}{(2\pi)^3}|F_2(k-q,q)|^2P_{\rm lin}(|k-q|)P_{\rm lin}(q)
\end{equation}
and we fix $\Sigma_s = 3 h^{-1}$Mpc, $\sigma_v = 1.9 h^{-1}$Mpc and $A_{\rm MC} = 0.05$ in all of our fits, as in \cite{alph}. The motivation for these choices is discussed in \cite{VM14}.

Given $P(k,\mu)$, we determine the multipole moments
\begin{equation}
P_{\ell}(k) = \frac{2\ell+1}{2}\int_{-1}^1 P(k,\mu)L_{\ell}(\mu),
\end{equation}
where $L_{\ell}(\mu)$ are Legendre polynomials. These are transformed to $\xi_{\ell}$ via
\begin{equation}
\xi_{\ell}(s) = \frac{i^{\ell}}{2\pi^2}\int dk k^2P_{\ell}(k)j_{\ell}(ks)
\end{equation}
We then use 
\begin{equation}
\xi(s,\mu) = \sum_{\ell}\xi_{\ell}(s)L_{\ell}(\mu) 
\end{equation}
and take averages over any given $\mu$ window to create any particular template:
\begin{equation}
\xi(s,\alpha_{\perp},\alpha_{||})_{F, {\rm mod}}(s) = \int_0^1d\mu F(\mu^{\prime})\xi(s^{\prime},\mu^{\prime}),
\end{equation}
where $\mu^{\prime} =  \mu\alpha_{||}/\sqrt{\mu^2\alpha_{||}^2+(1-\mu^2)\alpha_{\perp}^2}$ and $s^{\prime} = s\sqrt{\mu^2\alpha_{||}^2+(1-\mu^2)\alpha_{\perp}^2}$.

In practice, we fit for $\alpha_{\perp},\alpha_{||}$ using $\xi_0,\xi_2$ and $\xi_{W1}, \xi_{W2}$, where $W1$ and $W2$ represent transverse and radial wedges split at either $\mu_d = 0.5$ or $\mu_d = 0.64$. When fitting to Wedges, we fit to the data using the model
\begin{equation}
\xi_{W1, {\rm mod}}(s) = B_1\xi_{W1}(s,\alpha_{\perp},\alpha_{||}) + A_{W1}(s) 
\label{eq:xiw1mod}
\end{equation}
\begin{equation}
\xi_{W2, {\rm mod}}(s) = B_2\xi_{W2}(s,\alpha_{\perp},\alpha_{||}) + A_{W2}(s),  
\label{eq:xiw2mod}
\end{equation}
where $A_x(s) = a_{x,1}/s^2+a_{x,2}/s+a_{x,3}$.

To fit $\xi_0,\xi_2$, we recognize $\xi_2 = 5\int_0^1d\mu\left(1.5\mu^2\xi(\mu)-0.5\xi(\mu)\right)$ and, denoting $\int_0^1d\mu\mu^2\xi(\mu)$ as $\xi_{\mu2}$, we fit to the data using the model
\begin{equation}
\xi_{0, {\rm mod}}(s) = B_0\xi_{0}(s,\alpha_{\perp},\alpha_{||}) + A_{0}(s)  
\label{eq:xi0mod}
\end{equation}
\begin{equation}
\xi_{2, {\rm mod}}(s) = 5\left(1.5B_{\mu}\xi_{\mu2}(s,\alpha_{\perp},\alpha_{||})-0.5B_0\xi_0(s,\alpha_{\perp},\alpha_{||})\right) + A_{2}(s) 
\label{eq:xi2mod}
\end{equation}
For all $B_x$, the parameter essentially sets the size of the BAO feature in the template. We apply a Gaussian prior of width ${\rm log}(B_x) = 0.4$ around the best-fit $B_0$ in the range $45 < s < 80h^{-1}$Mpc with $A_x = 0$; this treatment assumes the amplitude of the BAO feature is isotropic. 
\section{Empirical Results}
\label{sec:BAOmeas}
\begin{table*}
\begin{minipage}{7in}
\caption{The statistics of BAO scale measurements recovered from the DR11 mock samples. `A14' results are taken from Anderson et al. (2014). All values are recovered from the distribution of the fits to the 600 mocks; $\langle\rangle$ denote the mean values, $S$ denotes standard deviation, and $C_{||,\perp}$ the denotes the correlation between the maximum likelihood values of $\alpha_{||}$, $\alpha_{\perp}$.   }
\begin{tabular}{lcccccccccccccccccccc}
\hline
\hline
Publication & Method  & $\langle \alpha_{\perp} \rangle$ &  $\langle \sigma_{\perp} \rangle$ & $S_{\perp}$ & $\langle \alpha_{||} \rangle$ &  $\langle \sigma_{||} \rangle$ & $S_{||}$ & $C_{||,\perp}$\\
\hline
Anderson et al. (2014) & M  & 0.9999 & 0.0137 & 0.0149 & 1.0032 & 0.0248 & 0.0266 & -\\
& W, $\mu_d=0.5$  & 0.9993 & 0.0161 & 0.0153 & 1.0006 & 0.0296 & 0.0264 &-\\
\hline
This work & M  & 0.9987 & 0.0150 & 0.0145 & 1.0017 & 0.0232 & 0.0257 & -0.49\\
& W, $\mu_d = 0.5$  & 0.9992 & 0.0159 & 0.0157 & 1.0010 & 0.0274 & 0.0274 & -0.56\\
& W, $\mu_d = 0.64$ &  0.9980 & 0.0153 & 0.0152 & 1.0032 & 0.0274 & 0.0276 & -0.54\\

\hline
\label{tab:mockbao}
\end{tabular}
\end{minipage}
\end{table*}

We use PTHalo \citep{Manera13} mock galaxy catalogs (mocks) to empirically test our our analytical derivations. The mocks we use were created to match the SDSS-III \citep{Eis11} data release 11 (DR11) BOSS \citep{Dawson12} CMASS sample. The imaging \citep{F,C} and spectroscopic data \citep{Smee13} were obtained using the SDSS telescope \citep{Gunn06} and reduced as described in \cite{Bolton12}

The DR11 CMASS sample contains galaxies with $b\sim2$ \citep{White11} distributed over 8500 deg$^2$ with $0.43 < z < 0.7$. The 600 PTHalo mocks created to match this sample are described in \cite{Manera13} and \cite{alph}. Results for Wedges and Multipoles fitting to these mocks have previously been published in \cite{alph}, and we use the same post-reconstruction pair-counts as in \cite{alph}. We bin $\xi(s)$ in $s$ bins of width 8$h^{-1}$Mpc, matching the fiducial choice of \cite{alph} that was determined optimal in \cite{Per14}. We calculate $\xi(s,\mu)$ in $\mu$ bins of width $0.01$ using the \cite{LS} method, modified for reconstruction \citep{Pad12},
\begin{equation}
\xi(s,\mu) =\frac{DD(s,\mu)-2DS(s,\mu)+SS(s,\mu)}{RR(s,\mu)},
\label{eq:xicalc}
\end{equation}
where $D$ is the reconstructed data points, $R$ is a set of points randomly sampling the angular and radial selection functions, and $S$ is a separate set of these random points whose positions have been shifted by the reconstruction according to the reconstructed density field \citep{Pad12}. We then determine the correlation function for any particular window over $\mu$ via 
\begin{equation}
\xi_F(s) = \sum^{100}_{i=1} 0.01\xi(s,\mu_i)F(\mu_i), 
\label{eq:xiell}
\end{equation}
where $\mu_i = 0.01i-0.005$.

\begin{figure}
\includegraphics[width=84mm]{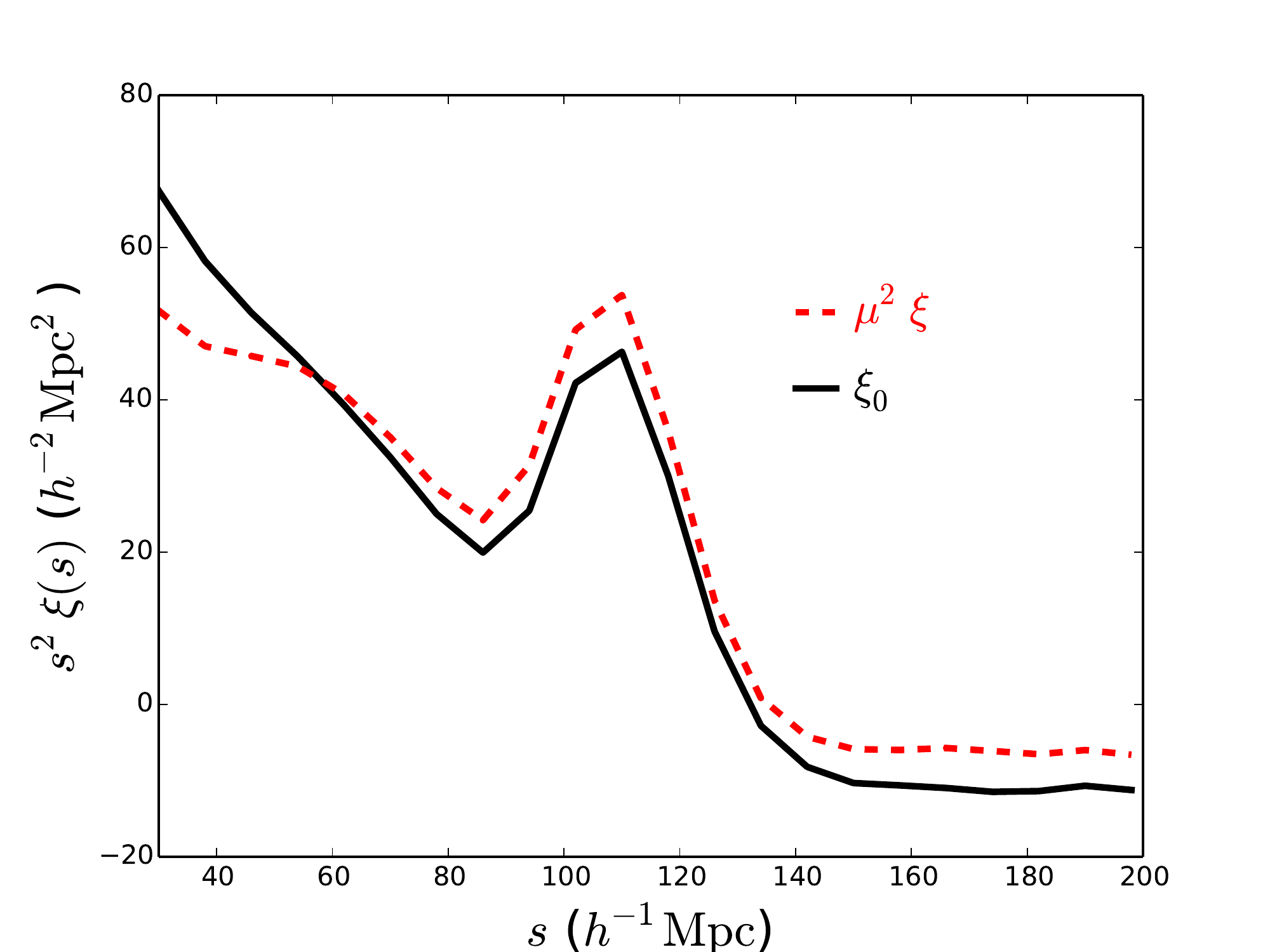}
  \caption{The mean $\xi_0$ and $3\int_0^1{\rm d}\mu\mu^2\xi(\mu)$, denoted $\mu^2\xi$, recovered from post-reconstruction DR11 CMASS mocks. }
  \label{fig:ximu2}
\end{figure}

Fig \ref{fig:ximu2} displays the mean $\xi_0$ recovered from these mocks post-reconstruction (black curve) compared to the mean $3\int_0^1{\rm d}\mu\mu^2\xi(\mu)$ moment (red curve). In principle, they should appear identical, as RSD have been removed in the reconstruction. However, differences are observed that are similar to the differences observed in post-reconstruction Wedges (see, e.g., figure 19 of \citealt{alph}).

We measure the line of sight, $\alpha_{||}$, and transverse, $\alpha_{\perp}$, BAO scale information for each of the 600 mocks using three different pairs of observables: \begin{enumerate}
  \item The combination of $\xi_0$ and $\xi_2$, as described by Eqs. \ref{eq:xi0mod} and \ref{eq:xi2mod}; we denote these results as `M' (for Multipoles)
  \item The combination of $\xi_{W1}$ and $\xi_{W2}$ Wedges split at $\mu_d = 0.5$; we denote these results as `W, $\mu_d=0.5$' 
  \item The combination of $\xi_{W1}$ and $\xi_{W2}$ Wedges split at $\mu_d = 0.64$; we denote these results as `W, $\mu_d=0.64$'
\end{enumerate}
For both Wedges, we use the model described by Eqs. \ref{eq:xiw1mod} and \ref{eq:xiw2mod}.

Our results are shown in Table \ref{tab:mockbao}, where we also display the results from \cite{alph}, denoted with `A14'. One can see that our implementation of Wedges split at $\mu_d = 0.5$ and Multipoles generally match closely with \cite{alph}, though variations of up to 10 per cent are found for some standard deviations and mean uncertainties.

The uncertainties and standard deviations are slightly worse than our analytic predictions, as can be seen by comparing the three left-hand columns to the three right-hand columns in Table \ref{tab:pre}. The discrepancies are greatest for $\alpha_{||}$ and for Multipoles; the recovered standard deviation on $\alpha_{||}$ is 14 per cent larger than expected for Multipoles, which is likely related to the fact that the correlation between $\alpha_{\perp}$ and $\alpha_{||}$ is 20 per cent larger than expected. Despite not matching our quantitative predictions, the Multipoles fits still match our qualitative predictions: they recover the smallest standard deviations, mean uncertainties, and correlation between $\alpha_{||}$ and $\alpha_{\perp}$.

The Wedges split at $\mu_d = 0.5$ produce the results closest to our analytic predictions; the recovered $\alpha_{\perp}$, $\alpha_{||}$, and their correlation are all between 3 and 5 per cent greater than predicted. We find that Wedges split at $\mu_d = 0.64$ results in only a small improvement in the variance of $\alpha_{\perp}$ and the correlation between $\alpha_{\perp}$ and $\alpha_{||}$ while producing a slight increase in the variance of $\alpha_{||}$. The $\mu_d = 0.5$ Wedges recover the least biased mean $\alpha_{\perp}$ and $\alpha_{||}$ of the three methods we apply, though the difference in the bias compared to the Multipoles results is negligibly small (at most 0.034$\sigma$).

\begin{figure}
\includegraphics[width=84mm]{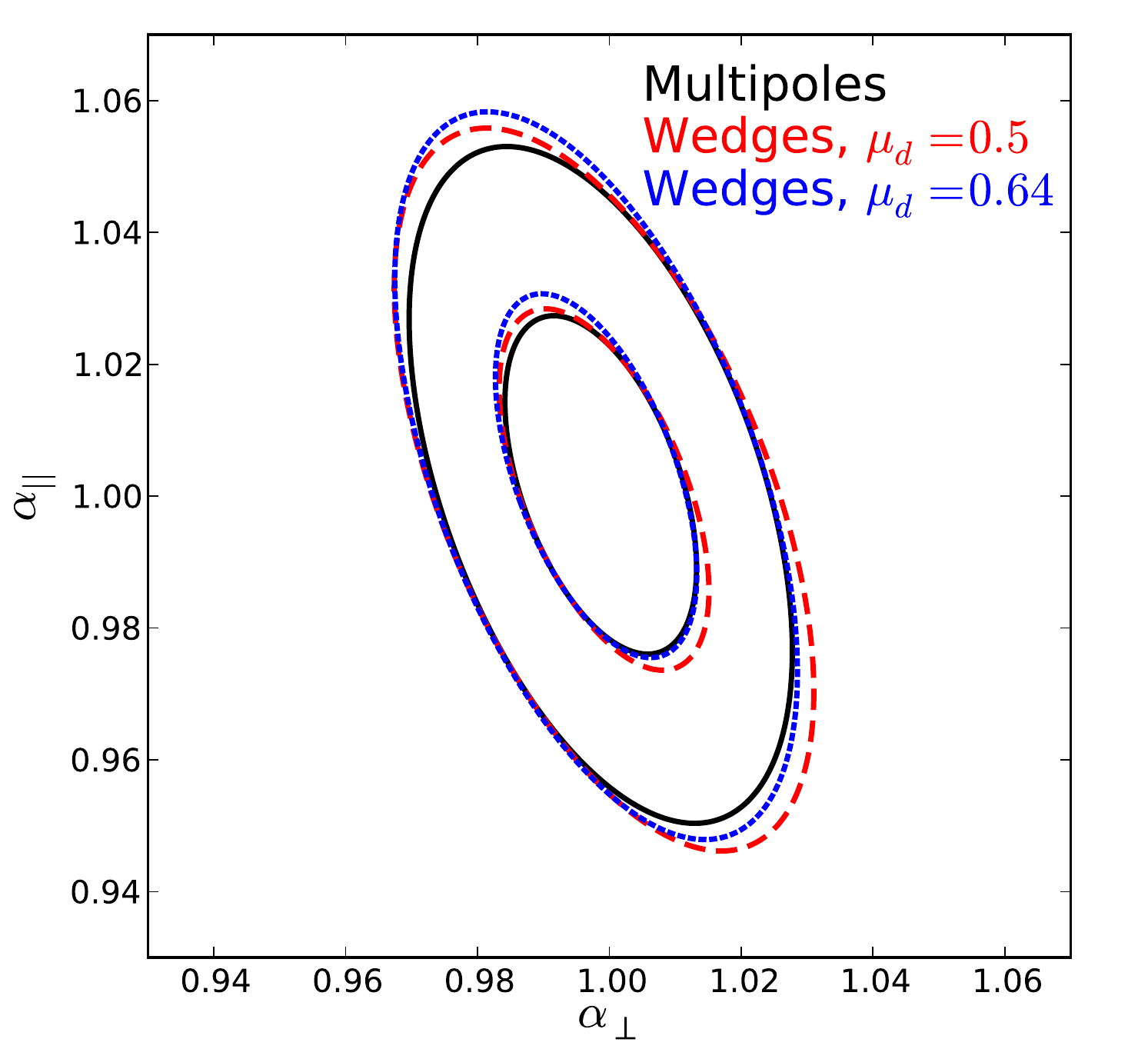}
  \caption{Ellipses showing the recovered standard deviation and correlations between the $\alpha_{||}$ and $\alpha_{\perp}$ for the different fitting techniques we apply, produced assuming these statistics describe a multivariate-Gaussian likelihood distribution.}
  \label{fig:ellipserecov}
\end{figure}

The results of our fits to the mocks are illustrated in Fig. \ref{fig:ellipserecov}, where we take the standard deviations and correlations of $\alpha_{||}$ and $\alpha_{\perp}$ for the different fitting techniques we apply and assume Gaussian statistics. Compared to Fig. \ref{fig:ellipsecom}, one can see that the ellipses are all more elongated (reflecting the increased uncertainty on $\alpha_{||}$ over those predicted). Similar to our predictions, the Multipoles ellipse is significantly smaller than the Wedges ellipses.

Table \ref{tab:corr} lists the correlations we find between the three different treatments we consider for $\alpha_{||}$ and $\alpha_{\perp}$, and the standard deviation obtained when averaging the measurements accounting for the correlation. The correlations are all greater than 0.85, and differences from 1 are caused mainly by the relative precision achieved in each method. Thus, there is negligible gain achieved by averaging the measurements, as one can see that there is at best a one per cent gain in the precision over that achieved by using only the Multipoles results.

\begin{table}
\caption{Correlations between the recovered $\alpha_{\perp}$ or $\alpha_{||}$ for different methods and the expected uncertainty when averages are taken incorporating the correlation. $C$ denotes correlation and $S_{C}$ denotes the standard deviation after combining the two measurements, accounting for the correlation.}
\begin{tabular}{lcccc}
\hline
\hline
Methods  &  $C_{||}$ & $C_{\perp}$ & $S_{C,||}$ & $S_{C,\perp}$ \\
\hline
Wedges $\mu_d = 0.5$; Multipoles & 0.85 & 0.86 & 0.0254 & 0.0144 \\
Wedges $\mu_d = 0.64$; Multipoles & 0.89 & 0.90 & 0.0256 & 0.0144\\
\hline
\label{tab:corr}
\end{tabular}
\end{table}

\section{Conclusions}

We have derived analytic formulae that describe the relative
importance of angular and radial dilation measured from moments of
2-point clustering statistics, with respect to the cosine of the angle
to the line of sight $\mu$. We have derived formulae for an arbitrary
window, $F$, that weights information with respect to $\mu$, and have
provided solutions for the cases where $F(\mu)$ is a power-law
(Section~\ref{sec:pfit}), a ``Wedge'' where $F(\mu)$ is 1 for given
range of $\mu$ and zero otherwise (Section~\ref{sec:wfit}), and where
the window is the 2nd-order Legendre polynomial (i.e., the clustering
observable is the quadrupole moment; Section~\ref{sec:qfit}). We have
presented results in real-space, valid for both the correlation
function, and in Fourier space for moments of the power
spectrum. These formulae extend the commonly used assumption that
isotropically averaged BAO provide a measurement of $D_V(z)$ to other
moments and allow for RSD when using power spectrum moments.

In Section~\ref{sec:err}, we derive the expected uncertainty of, and
covariance between, $\alpha_{\perp}$ and $\alpha_{||}$ obtained from a
pair of clustering measurements calculated for two different $F(\mu)$
assuming that information is evenly distributed in $\mu$ (as is
approximately the case for the BOSS CMASS galaxy sample). We show that
the optimal, maximum likelihood solution is the combination of the
monopole and quadrupole, or equivalently the monopole and $F(\mu) =
\mu^2$. We show that a third power-law window only adds degenerate
information and should not increase the statistical precision on
$D_A(z)$ and $H(z)$. We then find the optimal combination Wedges,
which we find are those split at $\mu_d=0.64$. For this optimal Wedge,
we predict the uncertainties on and correlations between $D_A(z)$ and
$H(z)$ are between 8 and 11 per cent larger than for the combination
of the monopole and quadrupole.

Our results differ from those of \cite{Taruya11,Kazin12}, as both studies found that including the hexadecapole significantly decreased the recovered uncertainty on $D_A(z)$ and $H(z)$. The key difference in our study is that we derive our results for post-reconstruction galaxy clustering measurements, where the Legendre polynomial moments are expected to be zero, except for the monopole. Thus, in our analytic formulation (supported by our empirical results), the inclusion of the quadrupole {\it does not} increase the total amount of BAO scale information (the covariance between the BAO information in the $p=2$ moment and in the monopole is the same as the variance expected for the $p=2$ moment), it simply allows for the information to be optimally projected into the $D_A(z),H(z)$ basis (and therefore {\it does} increase the total amount of cosmological information), and thus there is no additional information in the hexadecapole. In redshift-space, as studied by \cite{Taruya11,Kazin12}, the quadrupole and hexadecapole are expected to be non-zero and thus do contribute to the total amount of BAO information.

In our derivations, we consider only the $\mu$-dependent dilation at a
particular scale and assume the information at particular $\mu$ is
independent.  Such an assumption may be more appropriate in $k$-space,
where $P(k,\mu_1)$ and $P(k,\mu_2)$ are expected to be independent
(not accounting for any survey window function), but we test our
derivations using the redshift-space correlation function, where
$\xi(s,\mu_1)$ and $\xi(s,\mu_2)$ are not independent. Despite these
assumptions, the results we recover from test on mock samples closely
match our predictions, especially for $\alpha_{\perp}$, as presented
in Section \ref{sec:BAOmeas}.

Using the set of mock catalogues produced for the BOSS DR11 analysis,
we find that, as predicted, in terms of the recovered
uncertainty of, standard deviation of, and covariance between
$\alpha_{||}$ and $\alpha_{\perp}$, fitting to Multipoles
produces the optimal results of the three cases we test, matching our
analytic predictions. We also find, as predicted, Wedges split at $\mu_d = 0.64$ are optimal
compared to Wedges split at $\mu_d = 0.5$, although the decrease in
uncertainty is small ($< 5$ per cent). We find that the correlation between
Multipoles and Wedges is large enough that there is a negligible gain in information (~1 per cent
reduction in the standard deviation) when the results are combined.

We find a slight trend where the methods that depend most strongly on
clustering measurements at high $\mu$ are the most biased. The bias is
small, as the largest bias, found for the $\mu_d = 0.64$ Wedge, is
only 0.13$\sigma$. This trend is thus likely due to inaccuracies in
our modelling of the BAO feature at high $\mu$, where the non-linear
RSD signal is strongest. If the modelling as a function of $\mu$ can
be improved in future analyses, we expect the trend in bias will
decrease and that the recovered uncertainties and correlations will be
a closer match to our predictions for Multipoles. We therefore believe
that improving the $\mu$ dependence of the post-reconstruction BAO
template should be a priority for future BAO studies, and that by
doing so, the precision of the measurements made using Multipoles will
increase.

Our analysis provides further support for the future use of BAO to
make robust cosmological measurements. We have carefully considered
the meaning of BAO measurements made from moments of 2-point
functions, providing an optimal approach. Both this work, and the
recent work of \citet{Zhu14} who considered radial weighting of BAO
measurements, are testing and optimising the BAO measurement
methodology, increasing our understanding in line with the increasing
statistical precision afforded by future surveys. Our results, and the conclusions we draw, are specific to the case where information is evenly distributed in $\mu$. Thus, interesting possible
extensions include extending the methodology to more general cases
with different distributions of information with $\mu$ (e.g.,
Ly$\alpha$ or redshift-space measurements determined without using reconstruction), and allowing for correlations in $\mu$ in the covariance
matrix of $\xi({\bf r})$ required for small surveys. Such studies are likely to find that more than two moments are required to capture the full information content of the BAO signal.

\section*{Acknowledgements}

AJR acknowledges support from the University of Portsmouth and The
Ohio State University Center for Cosmology and AstroParticle
Physics. WJP acknowledges support from the UK STFC through the
consolidated grant ST/K0090X/1, and from the European Research Council
through the Darksurvey grant. We thank the anonymous referee, Daniel Eisenstein and Nikhil Padmanabhan for helpful
comments, Ariel Sanchez for comparisons with Wedges results, and Antonio Cuesta for providing all of the pair-counts we
used in our mocks analysis.

Mock catalog generation and BAO fitting made use of the facilities and
staff of the UK Sciama High Performance Computing cluster supported by
the ICG, SEPNet and the University of Portsmouth.

Funding for SDSS-III has been provided by the Alfred P. Sloan
Foundation, the Participating Institutions, the National Science
Foundation, and the U.S. Department of Energy Office of Science.
The SDSS-III web site is http://www.sdss3.org/.

SDSS-III is managed by the Astrophysical Research Consortium for the
Participating Institutions of the SDSS-III Collaboration including the
University of Arizona,
the Brazilian Participation Group,
Brookhaven National Laboratory,
Cambridge University ,
Carnegie Mellon University,
Case Western University,
University of Florida,
Fermilab,
the French Participation Group,
the German Participation Group,
Harvard University,
UC Irvine,
Instituto de Astrofisica de Andalucia,
Instituto de Astrofisica de Canarias,
Institucio Catalana de Recerca y Estudis Avancat, Barcelona,
Instituto de Fisica Corpuscular,
the Michigan State/Notre Dame/JINA Participation Group,
Johns Hopkins University,
Korean Institute for Advanced Study,
Lawrence Berkeley National Laboratory,
Max Planck Institute for Astrophysics,
Max Planck Institute for Extraterrestrial Physics,
New Mexico State University,
New York University,
Ohio State University,
Pennsylvania State University,
University of Pittsburgh,
University of Portsmouth,
Princeton University,
UC Santa Cruz,
the Spanish Participation Group,
Texas Christian University,
Trieste Astrophysical Observatory
University of Tokyo/IPMU,
University of Utah,
Vanderbilt University,
University of Virginia,
University of Washington,
University of Wisconson
and Yale University.

\label{lastpage}

\end{document}